\documentclass[prl,aps,showpacs,twocolumn,longbibliography,superscriptaddress, ]{revtex4-2}

\usepackage{color} 
\usepackage{bm}
\usepackage{dcolumn}
\usepackage{times}
\usepackage{amssymb} 
\usepackage{amsmath} 
\usepackage{braket}
\usepackage{ragged2e}
\usepackage{graphicx}
\usepackage{epstopdf}
\usepackage[english]{babel}
\usepackage{mathrsfs}
\usepackage{textcomp}
\usepackage{amsthm}
\usepackage{amsmath}
\usepackage{amsfonts}
\usepackage[dvipsnames]{xcolor}
\usepackage{soul}
\usepackage{braket} 
\usepackage{hyperref}
\usepackage{cleveref}
\usepackage{bibunits}
\defaultbibliography{HHGAAH.bib}
\defaultbibliographystyle{apsrev4-2}

\newcommand{\be}{\begin{equation}}
\newcommand{\ee}{\end{equation}}
\newcommand{\bea}{\begin{eqnarray}}
\newcommand{\eea}{\end{eqnarray}}
\newcommand{\ba}{\begin{array}}
\newcommand{\ea}{\end{array}}

\begin{document}
\begin{bibunit}

\title{Unraveling Multifractality and Mobility Edges in Quasiperiodic Aubry-André-Harper Chains through High-Harmonic Generation}
\author{Marlena Dziurawiec}
\affiliation{Institute of Theoretical Physics, Wroc{\l}aw University of Science and Technology, 50-370 Wroc{\l}aw, Poland}

\author{Jessica O. de Almeida}
\affiliation{ICFO - Institut de Ci\`encies Fot\`oniques, The Barcelona Institute of Science and Technology, 08860 Castelldefels (Barcelona), Spain}

\author{Mohit Lal Bera}
\affiliation{ICFO - Institut de Ci\`encies Fot\`oniques, The Barcelona Institute of Science and Technology, 08860 Castelldefels (Barcelona), Spain}

\author{Marcin Płodzień}
\affiliation{ICFO - Institut de Ci\`encies Fot\`oniques, The Barcelona Institute of Science and Technology, 08860 Castelldefels (Barcelona), Spain}

\author{Maciej M. Ma\'ska}
\affiliation{Institute of Theoretical Physics, Wroc{\l}aw University of Science and Technology, 50-370 Wroc{\l}aw, Poland}

\author{Maciej Lewenstein}
\affiliation{ICFO - Institut de Ci\`encies Fot\`oniques, The Barcelona Institute of Science and Technology, 08860 Castelldefels (Barcelona), Spain}
\affiliation{ICREA, Pg. Lluis Companys 23, ES-08010 Barcelona, Spain}

\author{Tobias Grass}
\affiliation{DIPC - Donostia International Physics Center, Paseo Manuel de Lardiz{\'a}bal 4, 20018 San Sebasti{\'a}n, Spain}
\affiliation{Ikerbasque - Basque Foundation for Science, Maria Diaz de Haro 3, 48013 Bilbao, Spain}
\affiliation{ICFO - Institut de Ci\`encies Fot\`oniques, The Barcelona Institute of Science and Technology, 08860 Castelldefels (Barcelona), Spain}

\author{Utso Bhattacharya}
\affiliation{ICFO - Institut de Ci\`encies Fot\`oniques, The Barcelona Institute of Science and Technology, 08860 Castelldefels (Barcelona), Spain}

\begin{abstract}
Quasicrystals are fascinating and important because of their unconventional atomic arrangements, which challenge traditional notions of crystalline structures. Unlike regular crystals, they lack translational symmetry and generate unique mechanical, thermal, and electrical properties, holding promise for numerous applications. In order to probe the electronic properties of quasicrystals, tools beyond linear response transport measurements are needed, since all spectral regions can be affected by the non-periodic geometry. Here we show that high-harmonic spectroscopy offers an advanced avenue for this goal. Focusing on the quasiperiodic 1D Aubry-André-Harper (AAH) model, we leverage high-harmonic spectroscopy to delve into their intricate characteristics: By carefully analyzing emitted harmonic intensities, we extract the multifractal spectrum -- an essential indicator of the spatial distribution of electronic states in quasicrystals. Additionally, we address the detection of mobility edges, vital energy thresholds that demarcate localized and extended eigenstates within generalized AAH models. The precise identification of these mobility edges sheds light on the metal-insulator transition and the behavior of electronic states near these boundaries. Merging high-harmonic spectroscopy with the AAH model provides a powerful framework for understanding the interplay between localization and extended states in quasicrystals for an extremely wide energy range not captured within linear response studies, thereby offering valuable insights for guiding future experimental investigations.
\end{abstract}

\maketitle


Investigating metal-insulator transitions is a fundamental inquiry within condensed matter physics. Anderson localization is a highly illustrative example of such transitions\cite{Anderson1958}, where a system becomes an insulator due to disorder. To explore this type of transition, it is common to study disordered non-interacting models that display Anderson localization. 
However, simple one-dimensional Anderson models do not exhibit metal-insulator transitions, as they remain insulators regardless of the disorder strength. Recently, quasi-periodic systems have garnered significant attention as an alternative means to study localization and criticality. These models, unlike periodic or disordered systems, reveal non-trivial localization properties even in one spatial dimension. A well-known example demonstrating a metal-insulator transition in one dimension is the Aubry-André-Harper (AAH) model \cite{AubryAndre1980,Harper1955}. This model can be viewed in terms of the superposition of two incommensurate lattices. When one of the lattices is treated as a weak perturbation, an incommensurate quasi-periodic potential emerges. The AAH model has been experimentally realized in various setups, including ultracold atoms in optical lattices \cite{L¨uschen2018,Modugno2010,Roati2008,Lohse2016,Nakajima2016,An2021} and photonic devices\cite{Tanese2014}, allowing researchers to gain deeper insight into the localization transition. 

For specific values of the quasi-periodic potential in the AAH model, a transition between ergodic and localized states occurs \cite{Tang1986,Hiramoto1989}.  In practice, this transition can be observed in transport experiments by studying the dynamics of particles or waves in these systems and measuring quantities like diffusion, conductivity, or a few transport exponents. However, at the critical point, the spatial distribution of states shows different degrees of localization or delocalization across the system, as such, the system's behavior exhibits a wide range of different scaling behaviors or scaling exponents at different spatial scales. This behavior at the critical point is also called multifractality:
While fractals are objects with a self-similar pattern, multifractals are objects with multiple patterns. Accordingly, fractals can be characterized by a scaling law with one non‐integer exponent, the fractal dimension, for a multifractal this is not enough to describe its dynamics; instead, a continuous spectrum of exponents, or singularity spectrum, is needed. Thus, determining multifractal behavior in quasiperiodic systems can be a very challenging task.

Moreover, according to scaling theory, in one and two dimensions, infinitesimal random disorder leads to the exponential localization of all single-particle states, resulting in the absence of diffusion \cite{Harper1955}. However, in three-dimensional Anderson systems, localized and extended states can coexist at different energies. The critical energy level known as the single-particle mobility edge (SPME) separates localized and extended eigenstates in the energy spectrum \cite{Harper1955,Tang1986}. Understanding the SPME is crucial in unraveling various fundamental phenomena, including metal-insulator transitions and thermoelectric response. Strikingly, quasiperiodic systems can manifest localization phase transitions and SPMEs even in one dimension (1D). 
However, due to the existence of a self-dual relation in the 1D Aubry-André-Harper (AAH) model, the localization lengths of states remain independent of energy, leading to the absence of an SPME. To generate SPMEs, short-range \cite{Hofstadter1976} or long-range hopping terms \cite{Madsen2013,Fraxanet2021,Fraxanet2022}, spin-orbit coupling \cite{Wang2016,Zeng2016}, or modified quasiperiodic potentials that violate the self-duality of the original AAH model \cite{Saha2019,Liu2017,Wang2016a,Yahyavi2019} can be included to form a generalized AAH model. However, the observation of SPME requires high-resolution measurements of electronic states across a wide energy range. Hence, experimental techniques with high sensitivity and resolution are essential.

In this Letter, we show that high harmonic spectroscopy can serve as a powerful tool to detect multifractal behavior, SPMEs, and related localization phenomena. High harmonic spectroscopy is a rapidly expanding field in strong-field attosecond science due to its potential to uncover the structural, topological, and dynamical properties of materials~\cite{Krausz2009,Calegari2016}. High-harmonic generation (HHG) is a nonlinear optical process resulting from the interaction between an intense laser field and a material~\cite{goulielmakis2022}, gas~\cite{constant1999,sutherland2004,li2020}, liquid~\cite{luu2018,neufeld2022,zeng2020} or crystal~\cite{ghimire2011}, producing high-order harmonics of the incident frequency. There has been a lot of interest in HHG spectra in liquids due to its statistical effects, where the dynamics exhibits dephasing and the energy levels have a multiplateu structure~\cite{zeng2020}. The same energy band structure is observed in the HHG of nonlocalized electrons in crystals~\cite{ndabashimiye2016,Li2017}. 
HHG allows for the generation of high-frequency harmonics, enabling probing of a broad energy range in the electronic band structure, including states near the mobility edge. Furthermore, the process of HHG is also sensitive to the electronic wave functions and their spatial distribution \cite{pattanayak2021}. As we will show, the nonlinear optical response can reveal information about the multifractal properties of the states.


Specifically, our study concentrates on  quasicrystals described by the AAH model from the point of view of high-harmonic spectroscopy. By analyzing the intensity of emitted harmonics, we demonstrate the potential to obtain the entire multifractal spectrum or distribution, which reveals crucial information about the localization properties of electronic states in these quasicrystals. Beyond the multifractal analysis, our research delves into the detection of mobility edges in generalized AAH models. We show that high-harmonic spectra can serve as a powerful tool for identifying critical energy points that separate localized and extended eigenstates. This provides valuable insights into the metal-insulator transition and the behavior of electronic states around the mobility edge. Through the combination of high-harmonic spectroscopy with the AAH model, our theoretical study sheds light on the intricate interplay between localization and extended states in quasicrystals.


The 1D Aubry-André-Harper (AAH) model~\cite{AubryAndre1980, Harper1955} model is given by the Hamiltonian
\small
\begin{equation}
    \label{eq:AAH}
    \hat{H} = -J\sum^N_{j=1} \left( c^{\dagger}_{j} c_{j+1} + c^{\dagger}_{j+1} c_{j}  \right) + 2\,V\sum^N_{j=1} \cos\left(2\pi\beta j \right) c^{\dagger}_j c_j,
\end{equation}
\normalsize
where $J$ is the hopping strength between nearest neighbors sites and $2V$ is the amplitude of the onsite potential. Parameter $\beta$, an irrational number modulating the lattice periodicity is chosen as the golden ratio $(\sqrt{5}-1)/2$, approximated by $\beta\approx\mathrm{Fibonacci}(n+1)/\mathrm{Fibonacci}(n)$ with the number of lattice sites $N=\mathrm{Fibonacci}(n)$. 
For the result presented here, we set $N=610=\mathrm{Fibonacci}(15)$.
The creation (annihilation) operators, $c^{\dagger}_{j}$ ($c_{j}$), create (annihilate) a spinless (or spin polarised) electron on lattice site $j$,
and the upper limit of the sum in the hopping term 
$N$ instead of $N-1$ implies periodic boundary conditions. 

 In a dominant onsite potential, $V>J$, the single particle electronic wavefunctions are localized i.e. concentrated to certain lattice sites and the material behaves as an insulator, as in the Anderson model with negligible contribution to charge transport. Whereas, when the nearest neighbor hopping is dominant, $J>V$, the electrons are delocalized, i.e.,
grossly uniformly distributed at all lattice sites and can exhibit charge transport in the presence of an infinitesimal electric field, thus behaving as a metal. 
By implementing a duality transformation of the form $c^{(\dagger)}_k = \frac{1}{\sqrt{N}}\sum_{j}e^{\pm i2\pi\beta k j} c^{(\dagger)}_j$ for the AAH model, a critical point, $V=J$, can be identified where the model is self-dual and exhibits multifractal properties~\cite{AubryAndre1980,Jose2018,Drese1997,hiramoto1992}. 
The effect of this transformation is to go to a momentum-space-like representation where wavefunctions that are delocalized in real space are localized in the dual space and vice versa. This is true except at the critical point where the wavefunctions are localized and delocalized simultaneously in both the spaces thus exhibiting multifractal behavior. 
 The duality relation also establishes a remarkable equivalence of energy levels in the AAH model:  For a given choice of parameter, $V/J = \alpha$, and for its inverse,  $V/J = \alpha^{-1}$, the energy levels are identical. Despite the identical energies, the single-particle states differ significantly, as these regions belong to distinct phases with opposite localization profiles.


We now couple the 1D Aubry-André-Harper (AAH) model with a strong linearly polarized incident laser pulse, the details of which are provided in the Supplemental Materials (SM) \ref{sm:model}. 
Our objective is to analyze the HHG spectrum, obtained via a windowed Fourier transform of the time-dependent lattice current operator (again, see SM \ref{sm:model} for details), and identify quantitative indicators of localized and delocalized phases within the AAH model. The HHG spectra carry crucial information about both the energy levels and the eigenstates of the system. 

\begin{figure}[ht!]
    \centering
    \includegraphics[width=1\columnwidth]{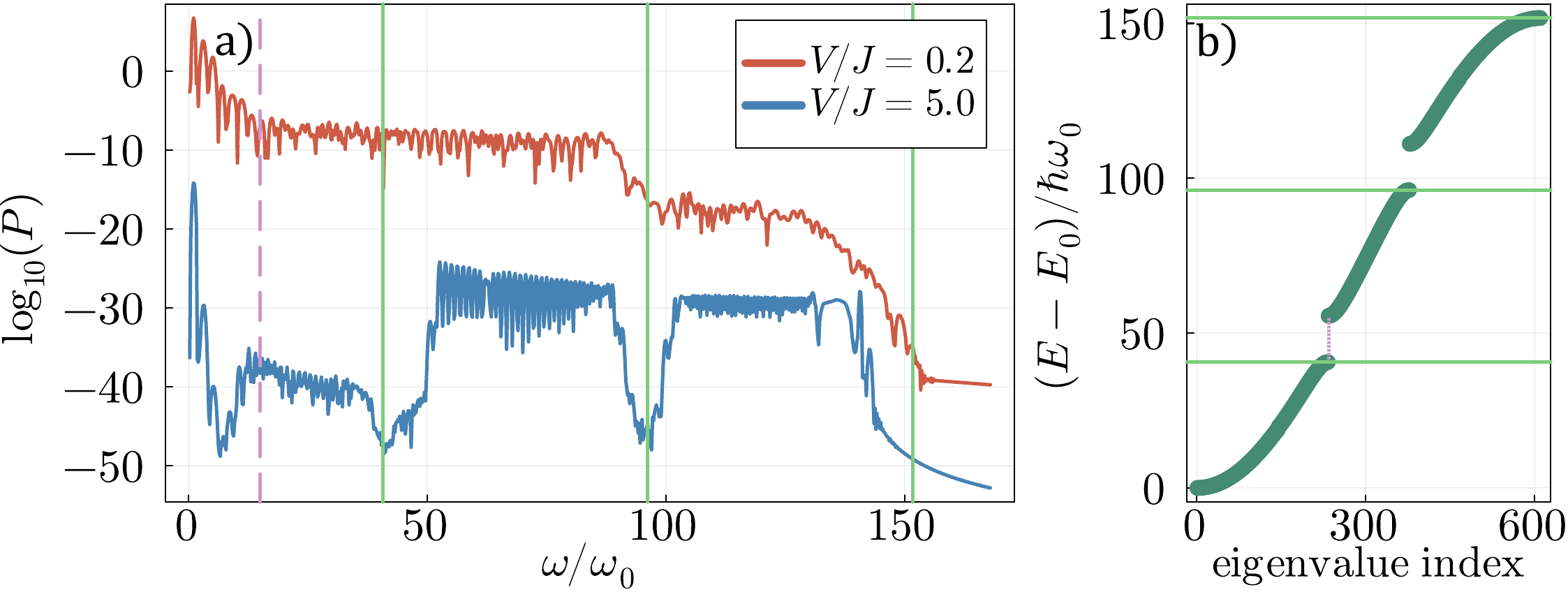}
    \caption{a) The HHG spectra for extended ($V/J = 5^{-1}$) and localized ($V/J = 5$) phases.  In the extended phase $J=0.15\approx 40\hbar\omega_0$ (red) and in the localized phase $V=0.15\approx 40\hbar\omega_0$ (blue). The system is half-filled. b) Corresponding energy spectra for both phases (overlapping).
    }
    \label{fig:HHGphases}
\end{figure}

In Fig.~\ref{fig:HHGphases}a, we present the high-harmonic emitted from the interaction of a pulse laser field with a system in the AAH model, for $V/J = 5^{-1}$ and $V/J = 5$. For comparison, Fig.~\ref{fig:HHGphases}b shows the energy spectrum which is equal for both phases. Despite the equal energy levels, the high harmonic spectra are crucially different, 
with a magnitude difference in the emission power for the two phases: The spectra of the extended phase (in red) have a metallic behavior with high emission power. In contrast, the localized phase (in blue) has features of insulating materials, with valleys in the energy bandwidths, as stressed with the solid green lines. This finding does not only serve as a clear signature of the contrasting localization behaviors, but importantly, it also highlights the high sensitivity of the high-harmonic spectrum to the spatial structure of the eigenstates. On the other hand, we note that in the AAH model, the character of the HHG spectrum is insensitive to the filling, as either all single particle states are completely localized or delocalized for a given value of the parameters of the system. This is in stark contrast to the HHG response from non-interacting semiconductors, where the HHG spectrum shows metallic or insulating behavior based on the partial or complete filling of a band. 

Although properties of the eigenstates are crucial for the high harmonic spectrum, there are also properties of the energy spectrum reflected in the high harmonic emission. Specifically, the pronounced valleys in the emission from the localized system coincide with the major gaps in the spectrum, displayed in the side panel of Fig.~\ref{fig:HHGphases}b. In both localized and extended cases, the HHG cutoff occurs at the maximum energy in the energy spectra. The first band gap is represented by the dashed purple line.

\begin{figure}[ht!]
    \centering
    \includegraphics[width=1\columnwidth]{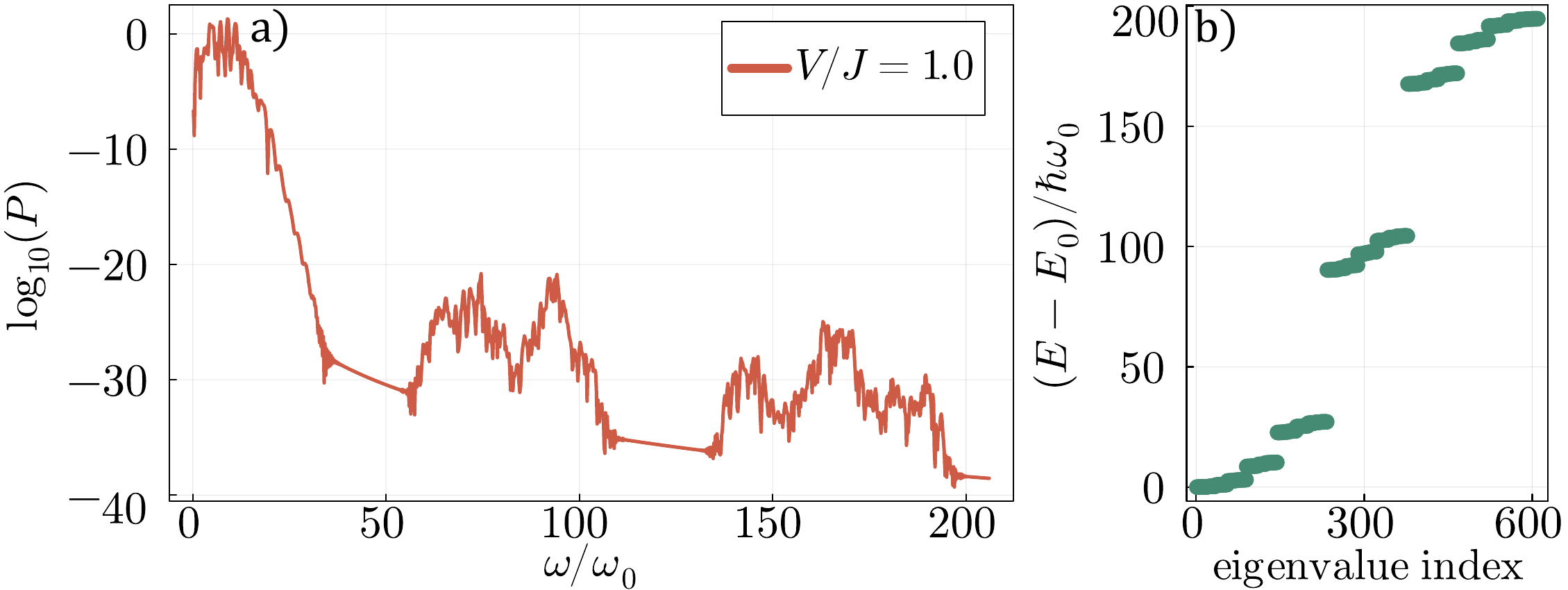}
    \caption{ a) The HHG spectrum and b) energy spectrum for the multifractal phase $V/J=1$. The system is half-filled.}
    \label{fig:Multifractal0}
\end{figure}

At the critical point, $V=J$, the AAH model is between the localized and delocalized phases and shows characteristic features common to both the phases, thus presenting multifractal behavior. We show the high harmonic and energy spectrum for the multifractal point
in Fig.~\ref{fig:Multifractal0}.
The HHG spectrum contains information about both the distribution of spectral gaps and about the degree of localization of eigenstates in the system. Here, we show that it is possible to read out the multifractal spectrum directly from the harmonic spectrum by using Multifractal Detrended Fluctuation Analysis (MDFA) and then contrast it against  the multifractal spectrum obtained via direct computation from the eigenstates. 



\begin{figure}[ht!]
    \centering
    \includegraphics[width=1\columnwidth]{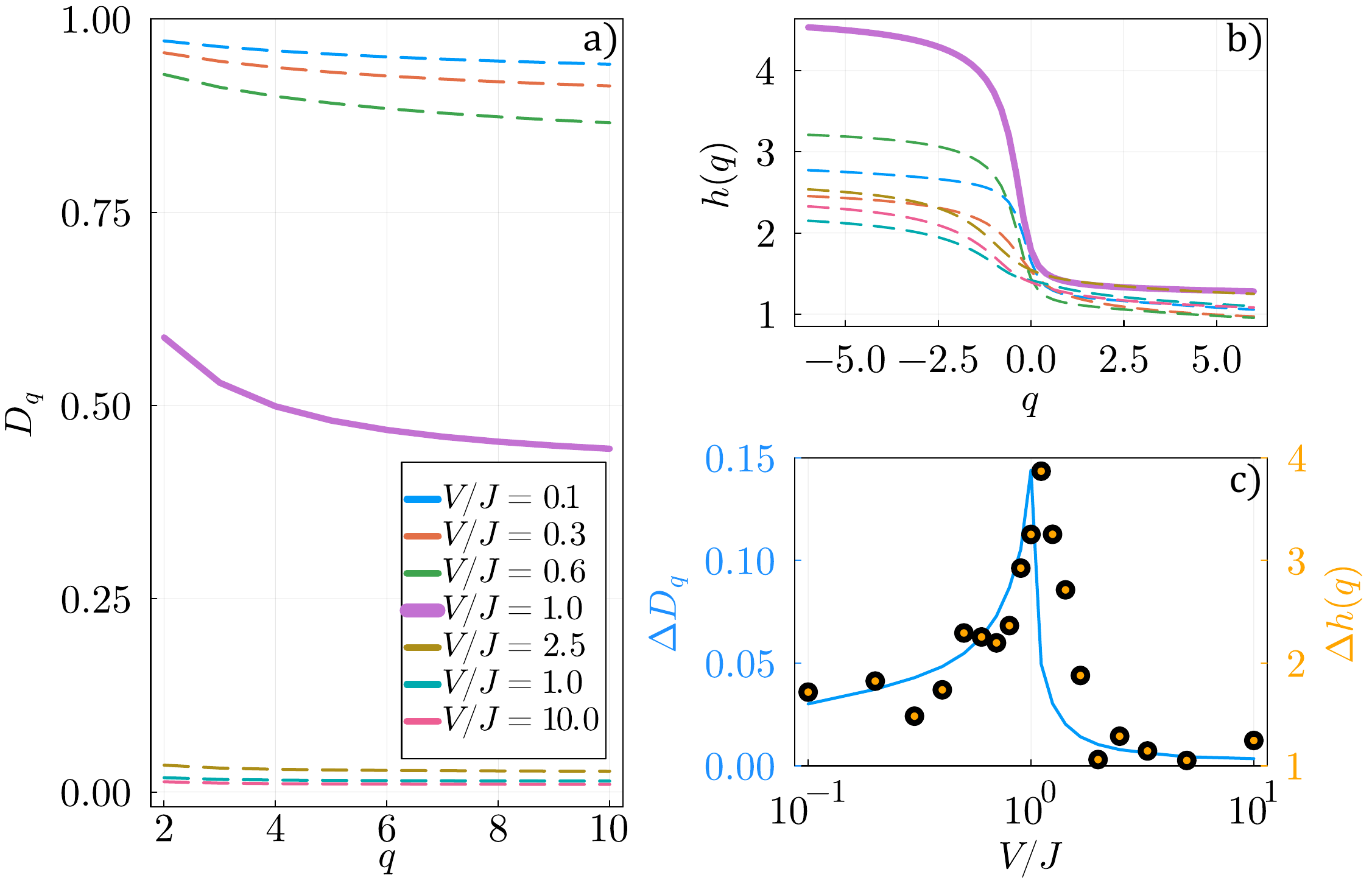}
    \caption{a) Fractal dimension as a function of $q$. b) Hurst exponent as a function of q. c) Width of curves $D_q$ (line) and $h(q)$ (dots) comparing a multifractality of the eigenstates and HHG spectra. 
    }
    \label{fig:Multifractal2}
\end{figure}

The fractal dimension quantifies the q-fractality of a quantum state. It involves partitioning the $\nu$-th eigenstate $|\Psi_{\nu}\rangle$ into $M$ segments of size $l$ and calculating generalized mean of cumulative probability in each segment,

\begin{equation}
    \label{eq:Dq}
    D_q(|\Psi_{\nu}\rangle)=\frac{1}{q-1} \frac{\log\left[\sum^{M-1}_{k=0}\left( \sum^l_{j=1} |\psi_{\nu,kl+j}|^2 \right)^q\right]}{\log (M^{-1})}.
\end{equation}
A completely localized state has dimension $D_q\sim0$, while a completely delocalized state has dimension $D_q\sim1$ and in both cases $D_q$ is non-dispersive. Figure~\ref{fig:Multifractal2}a shows the behavior of $D_q$ with $q$ for the different ratios $V/J$. At the critical point $V=J$, $D_q$ varies around $0.5$ and is dispersive with $q$, indicating multifractality. 
 Multifractals have multiple scaling exponents and their q-fractal dimension depends on the value of $q$ so $\Delta D_q = {\rm max}(D_q) - {\rm min}(D_q)$ is nonzero and maximal at the multifractal point, while it is near zero in the monofractal localized and delocalized phases as illustrated in Fig.~\ref{fig:Multifractal2}c.

We are now analyzing dynamical indicators of multifractality. Previous research \cite{Szabo2018} has concentrated on
the diffusion behavior in these systems. In the following, we will show that multifractality also manifests in the nonlinear  optical response. To this aim, we focus on the exponent $h(q)$ which is closely related to the generalized Hurst exponent. It typically represents autocorrelations within a time series, and quantifies how rapidly these correlations decline as the distance between value pairs increases. As explained in detail in the Supplementary Material (SM) section labeled "Multifractal Detrended Fluctuaction Analysis.", this exponent can be extracted from the power spectrum of HHG. We observe that for monofractals, the behavior of the dynamical $h(q)$ is close to constant, while for multifractals this behavior is rich. As shown in Fig.~\ref{fig:Multifractal2}b for different $V/J$, $h(q)$ is constant for both $q<0$ and $q>0$, with a gap at $q=0$.
The width of $h(q)$, $\Delta h(q) = { \rm max} [h(q)] - {\rm min} [h(q)]$, is higher than that of monofractals. Therefore, by quantifying the deviation of $h(q)$ from constant 
allows one to quantify multifractality. The dots in Fig.~\ref{fig:Multifractal2}c show that $\Delta h(q)$ has a peak for the values $V/J$ close to the critical point that closely resembles the behavior of $\Delta D_q$. 
This comparison clearly shows that the dynamical measure is interconnected with the q-fractal dimension $D_q$, which in fact can also be expressed as $\frac{q h(q)-1}{q-1}$. 
In summary, this demonstrates that the inherent static multifractal traits of the eigenstates can be decoded from the HHG spectrum using the MDFA technique.



We next turn to the study of mobility edges, i.e., critical energies that sharply separate localized and delocalized states for fixed system parameters. While the original AAH Hamiltonian does not exhibit a mobility edge at finite energy, such a feature appears already in simple generalizations of the AAH Hamiltonian. For concreteness, we choose
\small
\begin{equation}
    \label{eq:AAH}
    \hat{H} = \sum^N_{j=1} \left( J c^{\dagger}_{j} c_{j+1} + J^{*}c^{\dagger}_{j+1} c_{j}  \right) + 2\,V\sum^N_{j=1} \frac{\cos\left(2\pi\beta j \right)}{1-b\cos\left(2\pi\beta j \right)} c^{\dagger}_j c_j,
\end{equation}
\normalsize
where for any value of $b \neq 0$ the system shows a mobility edge, as this term breaks the self-duality relation in the model. Notably, there exists an analytical solution that yields the relation of the mobility edge with the hopping strength and onsite potential~\cite{Cai2023}. 

A key quantity that can identify mobility edges between localized and delocalized states is the inverse particpation ratio (IPR). The IPR of a $\nu$-th  eigenstate $\ket{\Psi_{\nu}} = \ket{\psi_{\nu,1}, \psi_{\nu,2}, \ldots, \psi_{\nu,L}}$ is defined as


\begin{equation}
    \mathrm{IPR}(\nu)=\sum\limits^{L}_{i=1}\lvert\psi_{\nu,i}\rvert^4.
\end{equation}

A zero IPR is a signature of delocalization, while a unit IPR represents localization. A sharp change in IPR indicates the presence of a mobility edge.
Subsequently, we show how the energy spectra and IPR of each eigenstate vary with $V/J$, for zero and non-zero $b$,
and how this behavior of IPR can also be captured directly in HHG. 

\begin{figure}[ht!]
    \centering
    \includegraphics[width=1\columnwidth]{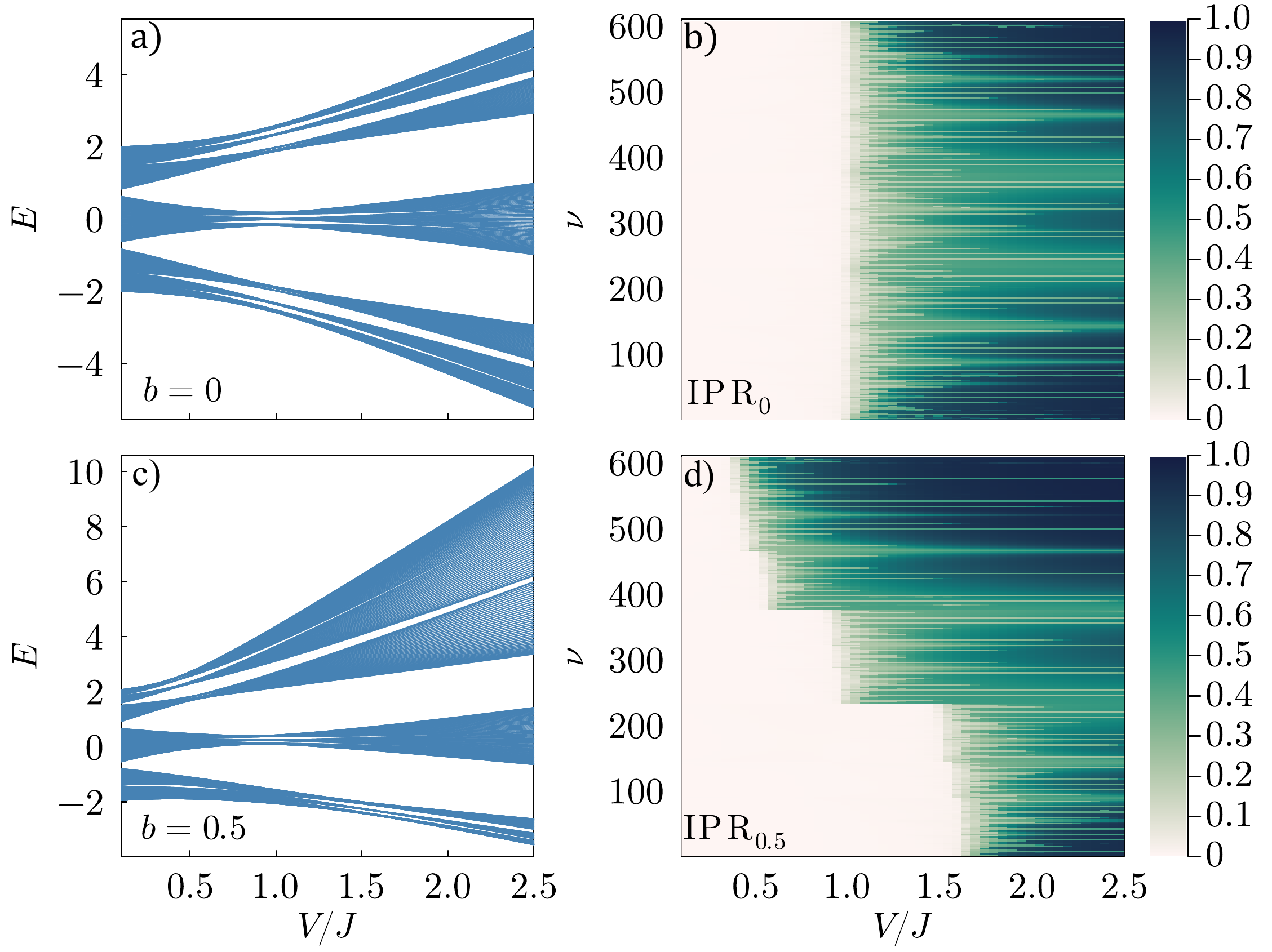}
    \caption{Energy levels as a function of $V/J$ in a) in the absence of SPME ($b=0$) and c) in the presence of SPME ($b=0.5$) and IPR for b) $b=0$ and d) $b=0.5$ of each eigenstate $\ket{\Psi_{\nu}}$ of the Hamiltonian \eqref{eq:AAH}.}
    \label{fig:SPME_IPR}
\end{figure}

\begin{figure}[ht!]
    \centering
    \includegraphics[width=1\columnwidth]{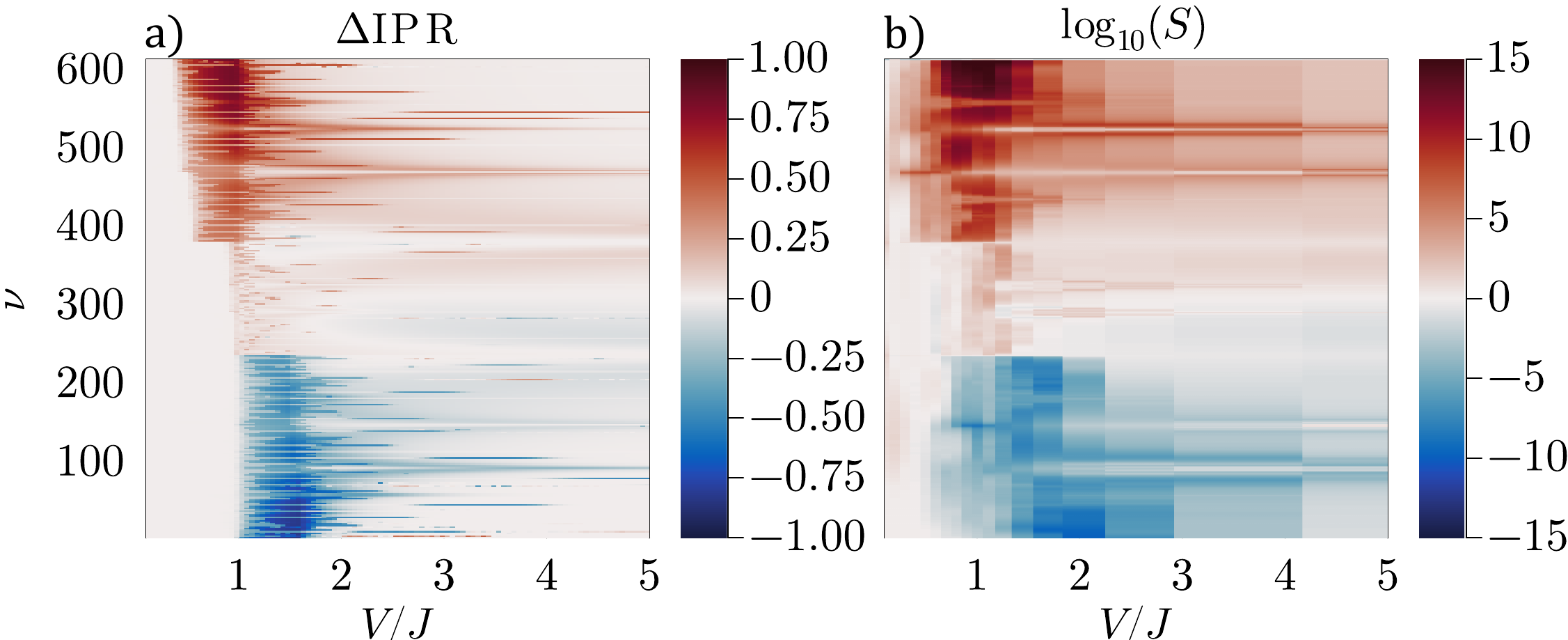}
    \caption{a) Difference in IPR with and without SPME for the different values of $V/J$ and eigenstate index $\nu$, as in Fig.~\ref{fig:SPME_IPR}. b) Ratio of below-bandgap contribution of HHG spectra \eqref{eq:S} as a function of filling $\nu$ and $V/J$.}
    \label{fig:SPME_S}
\end{figure}

We illustrate it in Fig.~\ref{fig:SPME_IPR}, where we present the
IPR versus $V/J$
for two cases when $b=0$ and $b=0.5$. We observe for $b=0$ a symmetric energy spectra and an absence of a mobility edge as is indicated by IPR that changes with filling (or energy) in an independent fashion and only at the critical point $V/J=1$,
separating states that are either all localized or all delocalized. For $b\neq0$ the energy spectra are asymmetric and the IPR shows a filling-dependent behavior typical of energy spectra with mobility edge. 
\begin{figure}[ht!]
    \centering
    \includegraphics[width=1\columnwidth]{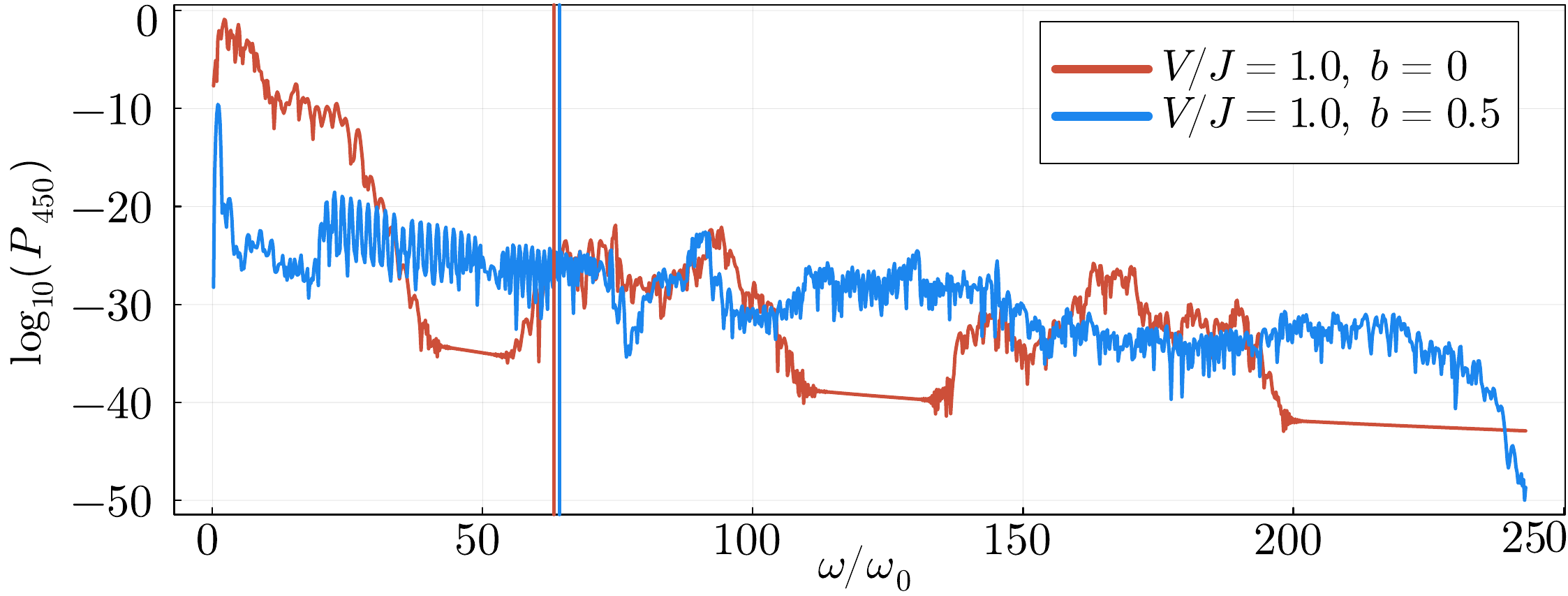}
    \caption{Example of the HHG spectra in the absence of SPME ($b=0$, red) and in the presence of SPME ($b=0.5$, blue). Vertical lines indicate the biggest gap between energy levels.}
    \label{fig:SPME}
\end{figure}
In order to examine, whether the HHG spectra can show a signature of mobility edges, we time-evolve the model
while changing the filling $\nu$ of the system in Eq.~\eqref{eq:current}. 
We compared the spectra in two cases, $b=0$ and $b=0.5$ for various fillings, see Fig.~\ref{fig:SPME} for example. For most fillings, in proximity of a critical point, the HHG spectra differ significantly.
As a rule of thumb, we find that the differences in the high harmonic spectrum originate from differences in the energy spectrum if the harmonics correspond to energies higher than the band gap of the system. However, contributions below the band gap strongly depend on the filling, which is connected with the existence of the mobility edges. To quantify this observation, we calculate the ratio between below-band gap contribution of spectra corresponding to systems with ($P_{0.5}$) and without ($P_0$) mobility edges,
\begin{equation}
\label{eq:S}
    S = \frac{\int_0^{\Delta E_0/\hbar} P_0 (\omega) d\omega}{\int_0^{\Delta E_{0.5}/\hbar} P_{0.5} (\omega)d\omega}.
\end{equation}
We compare this quantity with the difference of the IPR in these two cases,
\begin{equation}
    \Delta \rm{IPR} = \rm{IPR}_{0.5} -\rm{IPR}_{0}.
\end{equation}
The results are presented in Fig.~\ref{fig:SPME_S}. We notice that the below-gap HHG contribution indeed seems to be connected with the difference in IPR. For low fillings, for values of $V/J$ close to the mobility edge, when the eigenstates of Hamiltonian with SPME ($b=0.5$) are more extended than the eigenstates of Hamiltonian without them ($b=0$), the below-gap contribution is higher for the SPME Hamiltonian. For high fillings, the opposite is the case. In contrast, for moderate fillings or far from mobility edges, the below-gap contribution gets similar for both cases. Therefore, the presence of mobility edges can be clearly detected based solely on the high-harmonic spectra below the band gap. 

Characterization of the whole multifractal spectrum and detection of mobility edges pose significant experimental challenges for quasiperiodic systems. The results presented here theoretically addresses these problems through the lens of HHG and 
pave the way for future experimental investigations, contributing to a deeper understanding of the multifractality and mobility edges in quasicrystals and their potential applications in condensed matter physics.

\begin{acknowledgments}
T.G. acknowledges funding by BBVA Foundation (Beca Leonardo a Investigadores en Física 2023) and Gipuzkoa Provincial Council (QUAN-000021-01). U.B. acknowledges the project that gave rise to these results, received the support of a fellowship (funded from the European Union’s Horizon 2020 research and innovation programme under the Marie
Sklodowska-Curie grant agreement No 847648) from “la Caixa” Foundation (ID 100010434). The fellowship code is “LCF/BQ/PR23/11980043”. M.D. and M.M.M acknowledge support from the National Science Centre (Poland) under Grant No. DEC-2018/29/B/ST3/01892. M.P. acknowledges the support of the Polish National Agency for Academic Exchange, the Bekker programme no: PPN/BEK/2020/1/00317.
\end{acknowledgments}
\putbib
\end{bibunit}

\clearpage
\setcounter{equation}{0}
\setcounter{figure}{0}
\setcounter{table}{0}

\makeatletter
\renewcommand{\theequation}{S\arabic{equation}}
\renewcommand{\thefigure}{S\arabic{figure}}
\clearpage
\onecolumngrid
\begin{center}
    {\large\bf Supplemental Material for\\[2mm] ``Unraveling Multifractality and Mobility Edges in Quasiperiodic Aubry-André-Harper Chains\\[1mm]
    through High-Harmonic Generation''}\\[3mm]
    Marlena Dziurawiec,$^1$ Jessica O. de Almeida,$^2$ Mohit Lal Bera,$^2$ Marcin Płodzie\'n,$^2$\\
Maciej M. Ma\'ska,$^1$ Maciej Lewenstein,$^{2,3}$ Tobias Grass,$^{4, 5, 2}$ and Utso Bhattacharya$^2$\\[1mm]
{\small $^1${\it Institute of Theoretical Physics, Wroc{\l}aw University of Science and Technology, 50-370 Wroc{\l}aw, Poland}\\
$^2${\it ICFO - Institut de Ci\`encies Fot\`oniques, The Barcelona Institute of Science and Technology, 08860 Castelldefels (Barcelona), Spain}\\
$^3${\it ICREA, Pg. Lluis Companys 23, ES-08010 Barcelona, Spain}\\
$^4${\it DIPC - Donostia International Physics Center, Paseo Manuel de Lardiz{\'a}bal 4, 20018 San Sebasti{\'a}n, Spain}\\
$^5${\it Ikerbasque - Basque Foundation for Science, Maria Diaz de Haro 3, 48013 Bilbao, Spain}}
\end{center}
\setcounter{page}{1}
\vspace*{4mm}

\begin{bibunit}
\twocolumngrid
\subsection{Hamiltonian with Light}
\label{sm:model}
In the regime where the laser wavelength is much larger than the length of the system, the laser field coupling is well represented by the dipole approximation. The incident laser vector potential and electric field are:
\begin{equation}\label{E_A} \vec{A}(t) =
A(t)\hat{x},~~~~~\vec{E}(t) = -\partial_t\vec{A}(t).
\end{equation}
 
In the velocity gauge, the light-matter coupling induce a phase difference in the electronic hopping, dependent on the distance between neighbor sites, the Peierls phase, ${{\vec A}\cdot ({\vec r}_{n}-{\vec r}_{n'})}$
and therefore the time dependent hopping amplitude is,
\begin{equation}
    J(t) = J e^{i(n-n')A(t)} = J e^{i A(t)}.
\end{equation}
The distance is given in units of lattice constants and $A(t)$ is a time-dependent vector potential that describes the shape of the laser pulse,
\begin{equation}
A(t) = A_0 \sin^2\left(\frac{\omega_0 t}{2 n_c}\right) \sin\left(\omega_0 t\right),\quad 0<t<\frac{2\pi n_c}{\omega_0}.
\end{equation}
The number of cycles is $n_c = 5$, the laser field frequency is $\omega_0 = 0.004$ with amplitude $A_0 = 0.4$. 

The time-dependent Hamiltonian of the generalized AAH model is
\small
\begin{equation}
\label{eq:Ht}
    \hat{H}(t) =\sum^N_{j=1} \left(J(t) c^{\dagger}_{j} c_{j+1} + \mathrm{H.c.}\right) + 2\,V\sum^N_{j=1} \frac{\cos\left(2\pi\beta j \right)}{1-b\cos\left(2\pi\beta j \right)} c^{\dagger}_j c_j.
\end{equation}
\normalsize

The motion of the carriers within the bands creates a macroscopic current observable, with the current operator in the velocity gauge~\cite{Orlando2018,Jurss2019},
\begin{equation}
    \hat{I}(t) = i\sum_{j=1}^N \left(J(t) c^{\dagger}_{j} c_{j+1} - J^*(t) c^{\dagger}_{j+1} c_{j}  \right),
\end{equation}
and the expectation value is given by,
\begin{equation}
    \label{eq:current}
   I(t) =  \sum_{j=1}^{\nu} \bra{\psi_j(t)} \hat{I}(t) \ket{\psi_j(t)},
\end{equation}
where $\nu$ is the number of particles in the system (filling), and $\ket{\psi_j(t)}$ is the time evolution of each occupied single-particle energy state from $t = 0$. Using the Crank-Nicolson approximation, the time evolution of the wavefunction is calculated as follows,
\begin{align}  \label{eq:CN}
\ket{\Psi(t+\delta t)}&=\exp[-i\,\mathcal{H}(t)\delta t] \ket{\Psi(t)}\nonumber \\ 
&\approx \frac{1-i\,\mathcal{H}\left(t+\delta t/2\right)\delta t/2}{1+i\,\mathcal{H}\left(t+\delta t/2\right)\delta t/2} \ket{\Psi(t)},
\end{align}
with time-step $\delta t$ and initial state $\ket{\psi_j(t=0)}$ that is the $j$-th eigenvector of the Hamiltonian \eqref{eq:AAH}. 



The Fourier transform of the derivative of the time dependent current is proportional to the emitted radiation in the frequency domain
\begin{equation} \label{eq:power}
P(\omega) = \left| \mathrm{FFT}\left[\dot{I}(t)\right]\right|^2.
\end{equation}

\subsection{Multifractal Detrended Fluctuaction Analysis}
\label{sm:MFDFA}
Here, we briefly describe our calculations of Multifractal Detrended Fluctuaction Analysis (MF-DFA). For more details about the method, see the original paper~\cite{Kantelhardt}. We use the HHG spectra as our data series.

Firstly, we obtain the series profile -- we calculate a cumulative sum of a mean-centered data,
\begin{equation}
    y(\omega) = \int_0^{\omega} \left[ \log_{10}P(\omega') - \langle \log_{10} P \rangle \right] d\omega',
\end{equation}
where $\langle \log_{10} P \rangle$ is the arithmetic mean.
The curve $y(\omega)$ is then divided into segments (``boxes'') of size $s$, with $N_s$ being the total number of segments. In the next step, each $y_{\nu}(\omega$) (part of $y(\omega)$ that is inside the segment $\nu=1,2,\ldots,N_s$) is being fitted with a second-order polynomial $\overline{y}_{\nu} (\omega)$, so we calculate local trends. We remove the trends by subtracting the fitting from the respective data part $y_{\nu}(\omega)$, then we calculate the variance for each segment,
\begin{equation}
    F^2_{\nu} (s) = \frac{1}{s} \int_{(\nu-1)s}^{\nu s} \left[ y_{\nu}(\omega) - \overline{y}_{\nu} (\omega)\right]^2 d\omega.
\end{equation}
Next, we obtain the generalized mean (power mean) of a variance over all segments,
\begin{equation}
    F_q(s) =  \left\{ \frac{1}{N_s}  \sum^{N_s}_{\nu=1} \left[F^2_{\nu} (s)\right]^{q/2} \right\}^{1/q}.
\end{equation}
The calculations are repeated for various segments sizes $s$, with fixed exponent $q$. Then, the curve $F_q(s)$ is fitted with a power-law, $F_q(s) \propto s^{h(q)}$. The exponent $h(q)$ is a generalized exponent, closely related to the generalized Hurst exponents $H(q)$, $H(q)=h(q)-1$ for $h(q)>1$.




\putbib
\end{bibunit}
\end{document}